\documentclass[twocolumn]{article} 

\usepackage{arxiv}

\usepackage[utf8]{inputenc}
\usepackage[T1]{fontenc}
\usepackage{hyperref}
\usepackage{url}
\usepackage{booktabs}
\usepackage{amsfonts}
\usepackage{nicefrac}
\usepackage{microtype}
\usepackage{graphicx}
\usepackage[numbers, sort&compress]{natbib}
\usepackage{doi}
\usepackage{amsmath}
\usepackage{algorithm}
\usepackage{algorithmic}
\usepackage{newfloat}
\usepackage{listings}

\floatstyle{ruled}
\newfloat{listing}{tb}{lst}{}
\floatname{listing}{Listing}

\renewcommand{\headeright}{A Preprint}
\renewcommand{\undertitle}{A Preprint}
\renewcommand{\shorttitle}{Agentic LLM Framework for Mechanistic Enzyme Design}

\title{Beyond Protein Language Models: An Agentic LLM Framework for Mechanistic Enzyme Design}

\author{
    Bruno Jacob\textsuperscript{*},
    Khushbu Agarwal, Marcel Baer, Peter Rice, Simone Raugei \\
    Pacific Northwest National Laboratory\\
    Richland, WA
}

\begin{document}

% ======================================================================
% ABSTRACT BLOCK
% ======================================================================
\twocolumn[
  \begin{@twocolumnfalse}
    \maketitle
    
    \begin{abstract}
        We present Genie-CAT, a tool-augmented large-language-model (LLM) system designed to accelerate scientific hypothesis generation in protein design. Using metalloproteins ({\it e.g.}, ferredoxins) as a case study, Genie-CAT integrates four capabilities---literature-grounded reasoning through retrieval-augmented generation (RAG), structural parsing of Protein Data Bank files, electrostatic potential calculations, and machine-learning prediction of redox properties---into a unified agentic workflow. By coupling natural-language reasoning with data-driven and physics-based computation, the system generates mechanistically interpretable, testable hypotheses linking sequence, structure, and function. In proof-of-concept demonstrations, Genie-CAT autonomously identifies residue-level modifications near {[Fe--S]} clusters that affect redox tuning, reproducing expert-derived hypotheses in a fraction of the time. The framework highlights how AI agents combining language models with domain-specific tools can bridge symbolic reasoning and numerical simulation, transforming LLMs from conversational assistants into partners for computational discovery.
    \end{abstract}

    
    \vspace{1cm}
  \end{@twocolumnfalse}
]

{
  \renewcommand{\thefootnote}{\fnsymbol{footnote}}
  \footnotetext[1]{Corresponding author: \texttt{bruno.jacob@pnnl.gov}}
}

% ======================================================================
% INTRODUCTION
% ======================================================================
\section{Introduction}
The design of proteins with tailored properties is a central challenge in computational biology, which requires integration of biochemistry, structural biology, and machine learning to generate mechanistically grounded hypotheses. Although the emergence of protein language models (PLMs) and diffusion-based design frameworks for structural predictions has revolutionized the field \cite{winnifrith2024, ruffolo2024}. These statistical approaches remain limited in inferring structure-function relationships and their ability to design protein sequences that achieve specific biochemical functions \cite{kortemme2024, vu2023}. Contemporary design pipelines increasingly blend autoregressive sequence modeling, score-based diffusion, and multimodal priors with evolutionary or physics-informed constraints \cite{malbranke2023, zhou2024, kortemme2024}, yet achieving precision in functional design typically requires expert-guided interpretation of literature and structural data to identify changes in the amino acid sequence (mutations) that enhance activity, stability, or binding specificity.

Proteins that rely on metal-based redox cofactors pose a unique computational challenge because these cofactors act as nanoscale electrical components enabling controlled electron transfer in metabolism (Liu et al. 2014). Their behavior is governed by redox potential, which determines how readily a cofactor donates or accepts electrons. For iron–sulfur clusters—especially the widespread [4Fe–4S] units present across all domains of life—small changes to the surrounding protein environment can substantially shift redox potential and thereby alter catalytic activity (Ding, Nakai, and Gong 2022; Ferruz et al. 2023). These shifts arise from subtle electrostatic interactions that current statistical protein models cannot capture reliably, making explicit electrostatic and redox modeling essential for mechanistic protein design.

Recent agentic frameworks have begun to bridge this gap by combining machine learning with tool use and physics-based computation to expand the design space and improve reliability \cite{ghafarollahi2024, liu2024autope}. The AI Scientist family demonstrates how LLMs can generate ideas, write and run code, and iterate to produce complete manuscripts in computational domains \cite{lu2024aiscientist, yamada2025aiscientist_v2}, while translational efforts are integrating LLMs with laboratory automation and biofoundries to close the loop between design and wet-lab testing \cite{bran2023chemcrow}. However, these systems lack the domain-specific integration of literature, structural analysis, and physics-based modeling needed for protein design.

To address these limitations, we present Genie-CAT: an LLM-guided agentic system that integrates the interpretive power of large language models with quantitative physical modeling and domain-specific computation to generate hypotheses in protein design. Using proteins with iron--sulfur clusters as a case study, the system combines four core capabilities into a unified architecture:

\begin{enumerate}
    \item \textbf{Literature-guided reasoning:} A RAG capability for question answering and hypothesis generation, built by ingesting a publication corpus of one class of metalloproteins (hydrogenases).
    \item \textbf{Structural analysis:} Parsing of protein structures from Protein Data Bank (PDB)\cite{bermanPDB2000}  to extract residue-level properties, such as polarity and active-site environments.
    \item \textbf{Electrostatic potential calculations:} Physics-based computation of electrostatic landscapes to probe charge distributions relevant to catalysis.
    \item \textbf{Redox potential modeling:} Machine learning models trained to predict the impact of residue configurations on the redox properties of [Fe-S] clusters and related cofactors.
\end{enumerate}

Together, these components form a modular and extensible framework that can ground LLM-driven reasoning in biochemical, structural, and energetic evidence, producing interpretable and actionable hypotheses for downstream experimental validation.

% ======================================================================
% RELATED WORK
% ======================================================================
\section{Related Work}

Research relevant to our system sits at the intersection of machine learning for protein design, physics-based modeling of enzyme electrostatics and redox chemistry, and agentic large-language model systems for scientific discovery.

\subsection{Protein language models and machine learning for protein design}

The past decade has witnessed a fundamental transformation in computational protein science, driven by modern deep learning approaches that reframe protein modeling as a learnable mapping between sequence and structure \cite{ding2022, ferruz2023}. In recent work, ProteinMPNN revolutionized structure-conditional sequence generation by learning to predict amino acid sequences that fold into specified backbone geometries \cite{winnifrith2024}. Building on this foundation, RFdiffusion then extended generative capabilities to structure generation itself, enabling de novo design of protein folds through denoising diffusion processes \cite{ruffolo2024}. State-of-the-art systems now blend backbone-conditional design, diffusion methods, and multimodal priors with evolutionary or physics-informed constraints \cite{malbranke2023, zhou2024, kortemme2024}. 

However, these data-driven pipelines face fundamental limitations when designing for specific biochemical functions. The core challenge lies in interpretability and mechanistic control \cite{vu2023, winnifrith2024}. 
%Current models frequently operate as statistical black boxes, capturing broad sequence-structure relationships but providing limited insight into the molecular mechanisms underlying specific enzymatic activities \cite{vu2023, winnifrith2024}. 
This limitation has motivated the development of hybrid approaches that combine the pattern recognition capabilities of language models with physics-based modeling and domain-specific reasoning \cite{ghafarollahi2024}. However, for metalloenzymes, where the catalytic function depends on precise cofactor coordination, electrostatic fields, and redox chemistry, purely statistical approaches struggle to generate mechanistically informed design hypotheses. 

\subsection{Physics-based modeling of enzyme electrostatics and redox chemistry}

Electrostatic interactions play a fundamental role in enzyme catalysis, particularly for metalloenzymes, where precise charge distributions around cofactors determine catalytic efficiency and selectivity. 
Long-range electrostatic fields shape the potential energy landscape surrounding redox-active centers, modulating electron transfer driving force and directionality \cite{baker2001, gaughan2022, chen2022}. The anisotropic electrostatic potential around [Fe-S] clusters reflects the organization of polar and charged residues that fine-tune redox potentials and mediate proton-coupled electron transfer \cite{era2021, harris2016, dereli2025}. 

However, above approaches are continuum models which neglect local polarization and electronic coupling effects. This has driven the development of hybrid QM/MM techniques that treat metal clusters quantum mechanically while embedding the surrounding protein environment classically \cite{cheng2021}. Such approaches have successfully clarified how hydrogen-bonding networks around iron-sulfur clusters influence reduction potentials, revealing linear relationships between local electric fields and experimental redox potentials. 

\subsection{Retrieval-augmented methods and Agentic LLMs}

The integration of large language models with external knowledge sources and computational tools has emerged as a promising approach for scientific applications requiring both reasoning and domain expertise. Retrieval-Augmented Generation (RAG) demonstrated how grounding LLM outputs in external corpora can reduce hallucination and enable knowledge-intensive question answering \cite{lewis2020retrieval}. Building on this foundation, the ReAct framework showed how interleaving chain-of-thought reasoning with tool use improves interpretability and task success in multi-step problem-solving scenarios \cite{yao2022react}.

These capabilities have been extended to create increasingly sophisticated agentic systems for scientific discovery. Recent ambitious frameworks such as \emph{The AI Scientist} demonstrate end-to-end automation in which LLMs generate research ideas, write and execute code, and iterate to produce complete manuscripts in computational domains \cite{lu2024aiscientist, yamada2025aiscientist_v2}, while translational efforts are integrating LLMs with laboratory automation and biofoundries to close the loop between design and wet-lab testing \cite{bran2023chemcrow}.

However, existing agentic frameworks face significant limitations when applied to mechanistic protein design challenges. General-purpose systems such as The AI Scientist lack the domain-specific tools needed for structural analysis, electrostatic modeling, and redox chemistry calculations. Conversely, domain-focused systems like AutoProteinEngine primarily emphasize statistical model orchestration rather than integrating physics-based reasoning with literature knowledge. Most critically, current approaches do not systematically combine multi-modal evidence---literature insights, structural data, electrostatic calculations, and predictive models---into unified, interpretable hypotheses for experimental validation.

% ======================================================================
% METHODS
% ======================================================================
\section{Methods}
\begin{figure*}[t]
    \centering
    \includegraphics[width=0.85\textwidth]{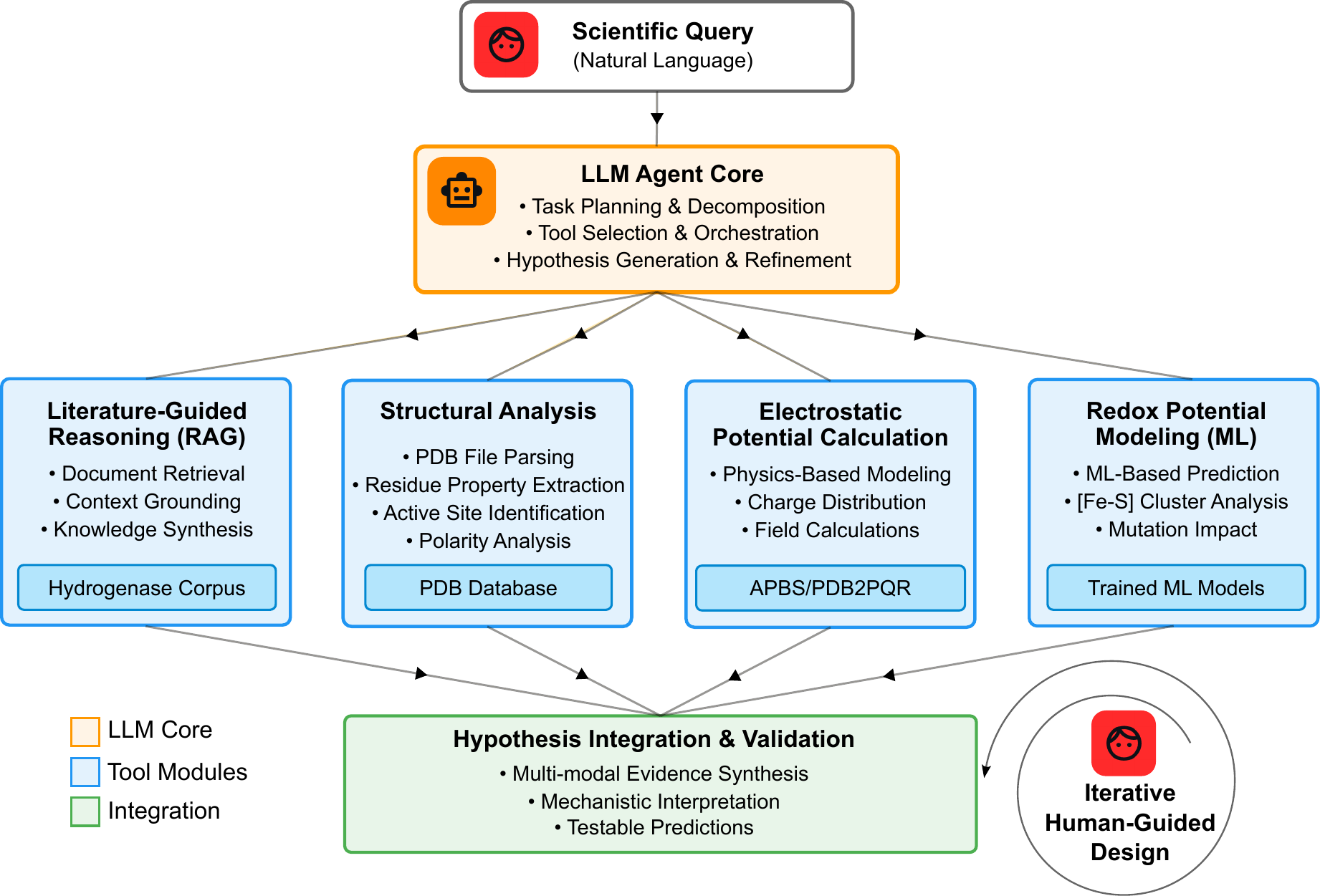}
    \caption{Overview of Genie-CAT's LLM-guided agentic system workflow: The system identifies the different components in the workflow to be executed based on user query.  Inputs (literature, PDB structures) are then processed through RAG, structural analysis, electrostatic calculations, and/or redox modeling. The evidence from different analysis is then summarized to the scientist enabling them to generate mechanistic hypotheses}
    \label{fig:system_overview}
\end{figure*}

Genie-CAT addresses these limitations by providing a modular architecture that specifically integrates the four types of evidence most critical for metalloenzyme design: literature-grounded reasoning through RAG, structural analysis of protein environments, physics-based electrostatic modeling, and machine learning prediction of redox properties. Rather than pursuing full experimental automation, our approach emphasizes generating mechanistically interpretable hypotheses that bridge the gap between statistical learning and physical understanding, positioning human experts to make informed design decisions based on multi-modal computational evidence. Figure ~\ref{fig:system_overview} provides an overview of the system, with the primary components discussed in following sections.

\subsection{Literature Ingestion and Indexing}

Our RAG design is primarily inspired by PaperQA2 \cite{skarlinski2024paperqa2}, where we augment each retrieved passage with a paper-level summary (e.g., the abstract or an offline precomputed document summary) to provide broader context in the prompt. Unlike PaperQA2's chunk-summary--based reranking, however, our summaries are used solely for contextualization at generation time alongside the retrieved chunk. This pairing of document-level summaries (global context) with chunk-level evidence (local context) allows us to rank the find the most relevant publications pertaining to the question while proving local context to answer the question correctly. Our approach can be viewed as a multiple-abstraction-level RAG design \cite{zheng2025multiple}.

To compile our literature corpus, we collected around 1600 relevant publications on hydrogenases and related metalloenzymes. 
%Source files were ingested as Portable Document Format (PDF) documents and converted to text with a layout-preserving extractor using pdfplumber with positional tolerances \cite{Singer-Vine_pdfplumber_2025}.  
The documents were parsed using pdfplumber library \cite{Singer-Vine_pdfplumber_2025}. The text was normalized and segmented into fixed-length overlapping windows to support context-aware retrieval. For this study, we adopted 500-character segments with 100-character overlap. Each segment is encoded with the \texttt{all-MiniLM-L6-v2} sentence transformer with 384-dimensional embeddings \cite{reimers2019sentence}.  The embeddings are stored in a FAISS index \cite{douze2024faiss} and configured for cosine similarity.

For evaluation, we follow the RAG Evaluation cookbook \cite{roucher2024rageval}. We synthesize 99 questions from the literature corpus using a generation and critique pipeline. An LLM generates candidate question–answer(QA) pairs, and critique agents score QA pairs on groundedness, relevance, and standalone clarity on a discrete 1 to 5 scale, discarding items with scores lower than 3. For answer grading, we reuse the cookbook’s LLM-as-a-judge prompt and its 1 to 5 correctness rubric. The scores reported in Table~\ref{tab:rag_metrics} are computed directly from these judge scores. Each question is answered in two isolated settings: (i) GPT-5-mini without retrieval (no RAG) and (ii) Genie-CAT with retrieval context with a medium thinking budget. Table~\ref{tab:rag_metrics} reports aggregated results over 10 runs: Genie-CAT with RAG achieves a higher mean score (4.38 vs. 4.01) and a 0.30 win rate with 0.55 ties, indicating consistent gains from retrieval while avoiding overconfident failures.

\begin{figure*}[h]
    \centering
    \includegraphics[width=0.9\textwidth]{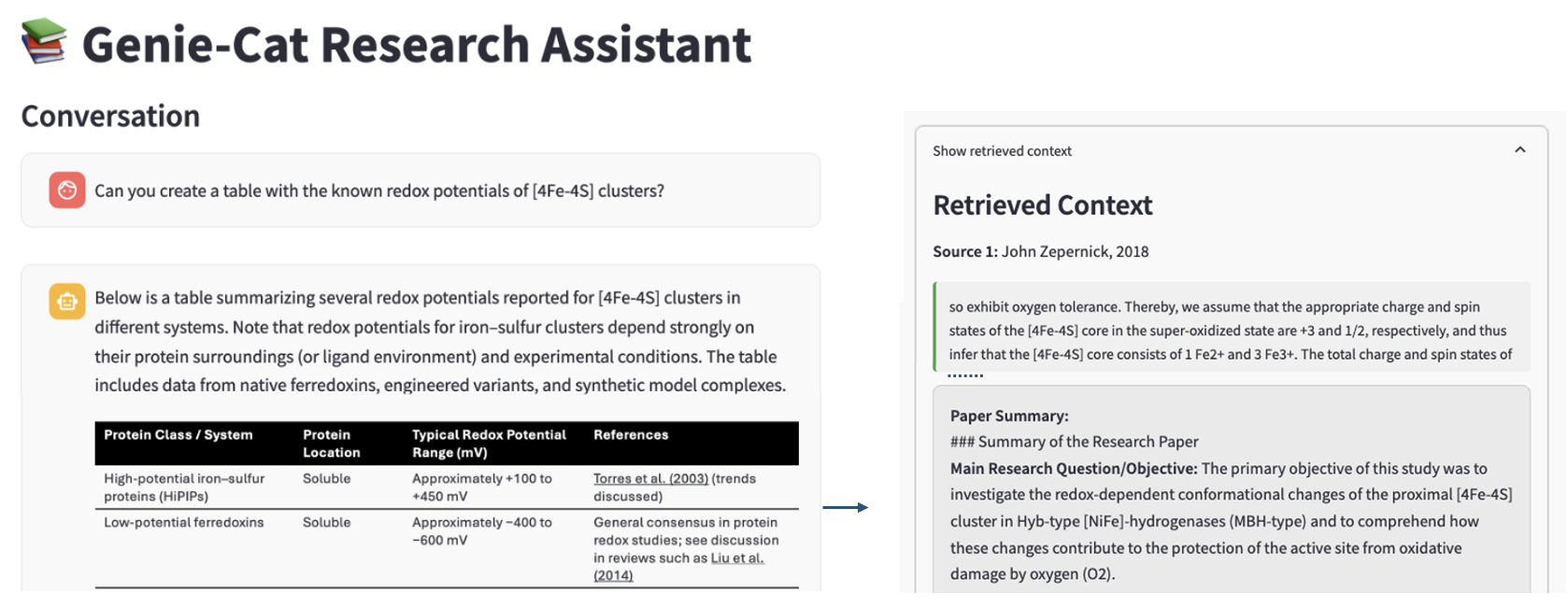}
    \caption{RAG Example query and response with retrieved context.}
    \label{fig:rag_example}
\end{figure*}

In addition to answering questions from literature, RAG component also allows scientists to click on citations to see the paper summary as well as retrieved sections of papers that were used to generate the answer. This enable trust and provides an opportunity for scientists to view the reasoning process.
Figure~\ref{fig:rag_example} shows an example of the user interface, RAG results and the view of the retrieved context. 

\begin{table}[t]
\centering
\caption{RAG evaluation using LLM-as-a-judge: Results are reported averaged over 10 independent runs. The system was given 99 questions to answer and an independent LLM was used to rate answer correctness using a scale of 1 (not correct)--5 (completely correct).}
\label{tab:rag_metrics}
\begin{tabular}{lcc}
\hline\hline
\textbf{Metric} & \textbf{GPT-5-mini (no RAG)} & \textbf{Genie-CAT} \\
\hline\hline
Mean (CI) & 4.01 (3.95--4.07) & \textbf{4.38 (4.35--4.42)} \\
Std & 0.09 & \textbf{0.05} \\
Min--Max & 3.89--4.14 & \textbf{4.30--4.47} \\
Win rate & 0.15 & \textbf{0.30} \\
Tie rate & \multicolumn{2}{c}{0.55} \\
\hline
\end{tabular}
\end{table}

\subsection{Structural Preprocessing and Residue-Level Analysis}
\begin{figure}[h]
    \centering
    \includegraphics[width=0.48\textwidth]{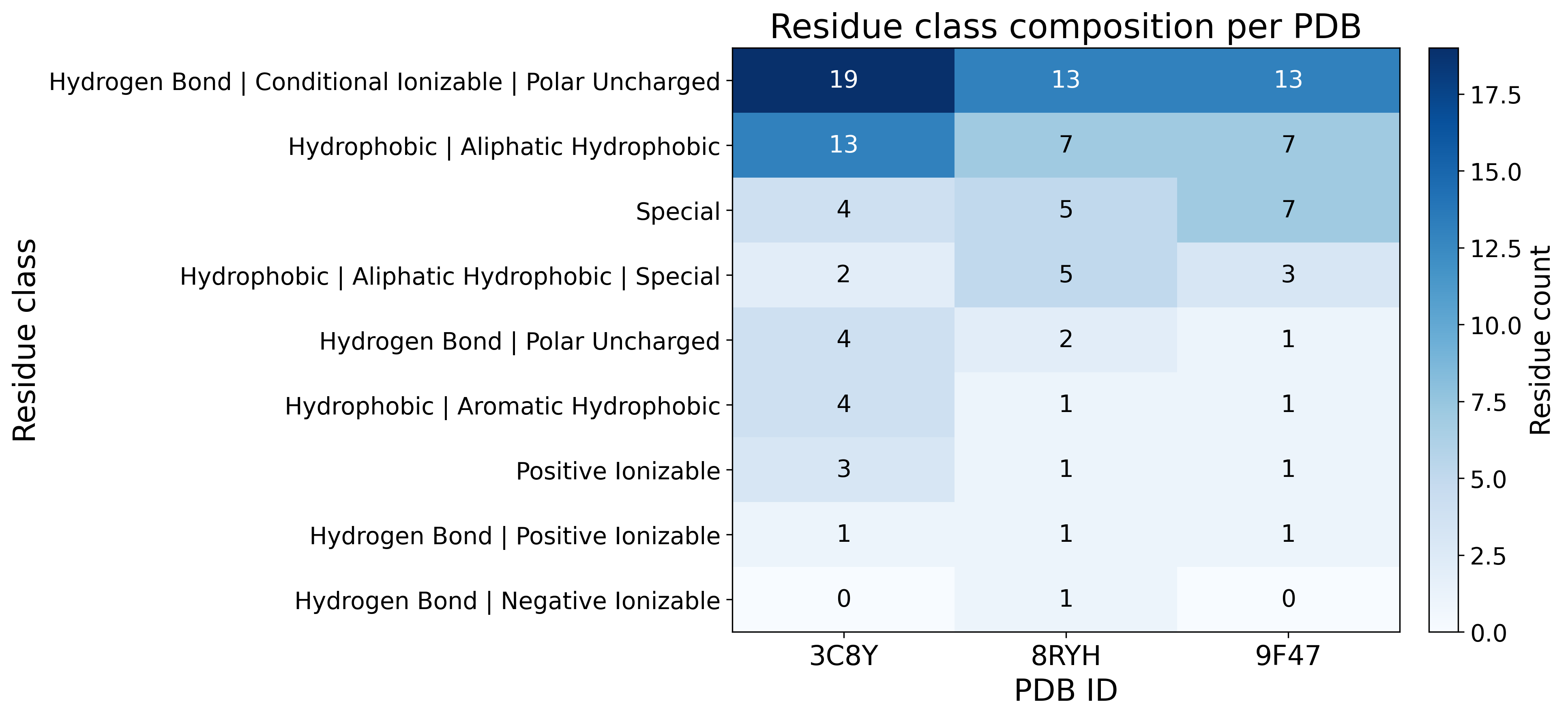}
    \caption{Example PDB structure analysis: Agent analyses PDBs to determine residue compositions, maps each residue to associated polarities and summarizes results}
    \label{fig:pdb_annotation}
\end{figure}

Genie-CAT provides functionality to parse protein structures available in Protein Data Bank (PDB) and extracting residue-level features such as polarity, coordination, and active-site environments.
Protein structures are accepted from three sources: user uploads, preloaded entries, and previously downloaded files. If none is found, the structure is automatically downloaded from the RCSB Protein Data Bank when a valid identifier is provided. Structures are parsed with MDAnalysis \cite{gowers2019mdanalysis,michaud2011mdanalysis} tool. Iron atoms are identified and used as reference points to compute distances from the centers of mass of the residues. The local environment is defined by a tunable spherical cutoff $R_{\rm cut}$. Each residue is assigned to physicochemical classes and recorded with its chain identifier and residue number. The procedure yields a summary of residue-level measurements and statistics. Genie-CAT then produces figures (distance histograms, class distributions, and chain breakdowns) using Matplotlib integration (Figure ~\ref{fig:pdb_annotation}) allowing scientists to study and compare-contrast protein structures and the associated properties.

%-- SR 2025-10-14, start
\subsection{Electrostatic Potential Computation and Visualization}
Electrostatic calculations are based on solving the Poisson-Boltzmann (PB) equation for a discrete charge distribution that represents the electrostatic field of a protein. We employ a three-stage workflow: structural preparation, APBS-based potential-grid computation, and evaluation of potentials on molecular surfaces.  

First, the protein structure is prepared for electrostatic calculations by identifying cofactors and their ligands, assigning point charges to each atom, and then evaluating the electrostatic potential. The point charges are taken from standard molecular mechanics force fields \cite{tian_ff19sb_2020}, literature, in-house databases, or estimated on the fly if they are not otherwise available. Our in-house database includes parameters for a variety of iron--sulfur cofactors (e.g., SF4 for cubanes, FES for di-iron motifs, F3S for open cubanes, etc.). Next, the PB equation is solved using the grid-based Adaptive Poisson-Boltzmann Solver (APBS)~\cite{jurrus_improvements_2018}. Finally, the electrostatic potentials are mapped onto molecular surfaces for further analysis and visualization. Visualization is performed by generating a PyMOL script that renders either surface-mapped electrostatic potentials or focused views of iron--sulfur clusters. Importantly, the agent is also provided with in-house tools for comparing the electrostatic maps of iron--sulfur cofactors across different proteins and for detecting changes due to protein modifications.
%-- SR 2025-10-14, end

\subsection{Redox Potential Model for Reasoning about Ligand Mutations}

The computed electrostatic field of the protein, is then used as input to predict the impact of residue changes and mutations on the redox properties of [Fe-S] clusters. We developed a novel machine learning model that enforces invariances reflecting the approximately $D_{2d}$-symmetric [4Fe-4S] core while accommodating protein-induced distortions. Concretely, the representation is invariant to global rotations and to within-cluster permutations (Fe and S subgroups), which regularizes learning and focuses the model on physically meaningful descriptors rather than orientation or atom ordering. This allows us to predict the effect of structural changes and amino acid substitutions on the redox potential rather than the absolute positioning of a ligand relative to Fe--S clusters. 

The redox potential model begins by encoding each Fe--S complex into a feature vector consisting of the positions of the atoms and the corresponding electric fields on each atom of the [Fe-S] clusters. From these quantities, we construct scalar features (e.g., sorted ligand descriptors, radial distances, pairwise cosines between bond vectors, and normalized triple products) that are permutation-invariant across the four atoms of each group; concatenating these sets yields an invariant encoding
$\phi_{g}\!\big(s_{i,g}, v_{i,g}\big)\in\mathbb{R}^{18}$, which is augmented with global electrostatic descriptors $Q_i$ (the electrostatic potential at the cluster center), $C_i$ (the electric field vector at the cluster center), and $\|C_i\|_2$:

\begin{equation}
\begin{array}{l}
    x_i = \Big[\,\phi_{\mathrm{Fe}}\!\big(s_{i,\mathrm{Fe}}, v_{i,\mathrm{Fe}}\big),\; \phi_{S_{1}}\!\big(s_{i,S_{1}}, v_{i,S_{1}}\big),\; \phi_{S_{2}}\!\big(s_{i,S_{2}}, v_{i,S_{2}}\big),\\[2pt]
\qquad Q_i,\; C_i,\; \|C_i\|_2\,\Big] \in \mathbb{R}^{57}.
\end{array}
\end{equation}

The redox potential is predicted by a fully connected multi-layer perceptron (MLP) $f_{\theta}:\mathbb{R}^{57}\rightarrow\mathbb{R}$ with two hidden layers of widths $(256,128)$, rectified linear activations, and a dropout layer with rate $0.1$ before the final linear layer, where $\theta$ denotes the network parameters. Before training, each component of $x_i$ and the target potential $E_i$ is standardized using training-set statistics so that the network operates on z-scored features while preserving the analytically enforced invariances. The normalized prediction for complex $i$ is $\hat{E}_i = f_{\theta}(\tilde{x}_i)$ with $\tilde{x}_i$ denoting the standardized feature vector.

Model parameters are optimized with AdamW \cite{loshchilov2017decoupled} using a learning rate of $10^{-3}$, decoupled weight decay coefficient $\lambda = 10^{-4}$, and minibatches of size $1024$. For a full dataset $\mathcal{D}$, split into subsets $\mathcal{D}_{\mathrm{train}}$ and $\mathcal{D}_{\mathrm{val}}$, we minimize the mean-squared error.
\begin{equation}
    \mathcal{L}(\theta) = \frac{1}{\lvert\mathcal{D}_{\mathrm{train}}\rvert} \sum_{i\in\mathcal{D}_{\mathrm{train}}} \bigl(E_i - \hat{E}_i\bigr)^2,
\end{equation}
where $E_i$ is the target redox potential and $\hat{E}_i = f_{\theta}(\tilde{x}_i)$ denotes the model prediction for sample $i$. Early stopping monitors the validation loss and retains the parameters attaining the minimum held-out error.

% ======================================================================
% SOFTWARE ARCHITECTURE
% ======================================================================
\section{Software Architecture}
Genie-CAT is implemented as a modular, tool-augmented language model system within a single containerized runtime. A Streamlit application provides the user interface and hosts a LangGraph ReAct agent that invokes typed tools for literature retrieval, structural analysis of Protein Data Bank files, and electrostatic potential computation and visualization in response to user queries (Figure~\ref{fig:system_overview}).
  
\noindent{\textbf{Modular, Extensible Design}.}
Each capability is implemented as an independent tool, enabling new physics engines or ML models to be integrated without modifying the core agent. This design supports planned extensions to high-fidelity simulations—including QM/MM, DFT cluster models, and large-scale MD—by allowing new tools to register input schemas, submit compute jobs, and return results through the same interface.

\noindent{\textbf{ReAct-based reasoning}.}
The agent uses a ReAct (Reasoning and Acting) \cite{yao2022react} pattern: the LLM iteratively generates thoughts, invokes tools (RAG retrieval, ESP calculations, redox prediction, Shell, Python REPL), observes results, and synthesizes answers grounded in retrieved or computed evidence. Tool selection and ordering are determined dynamically based on the query and intermediate results.
  
\noindent{\textbf{Performance Characteristics}.}
Typical queries require 1–5 LLM calls depending on complexity. RAG-only queries and structural parsing queries respond in ~1–5 seconds, while  ESP computations complete in ~120-180 seconds.  The average inference time for redox model was around 20 seconds. Future HPC integration will allow long-running simulations (QM/MM, extended MD) to run asynchronously, with the system returning a job ID and later fetching results into the reasoning chain.

% ======================================================================
% CASE STUDY
% ======================================================================
\section{Agentic Multi-Modal Reasoning on Metalloenzymes: A Ferredoxin as a Test Case}
\begin{figure*}[h]
    \centering
    \includegraphics[width=0.86\textwidth]{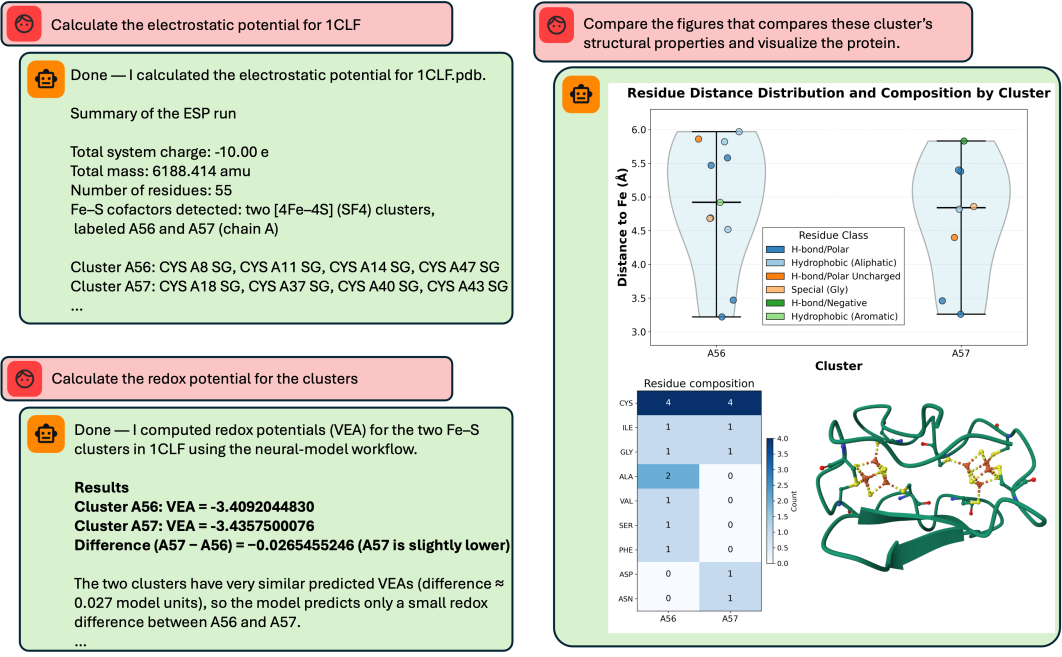}
    \caption{Example interactive session with Genie-CAT on ferredoxin 1CLF. The user issues three sequential queries: (Q1) compute electrostatic potentials for 1CLF, (Q2) predict the redox potentials of its two [4Fe–4S] clusters, and (Q3) generate comparative figures and tables, including a structural plot of the protein. For each query, the agent selects and orchestrates the appropriate tools (APBS-based electrostatics, symmetry-aware redox predictor, and plotting utilities), then synthesizes the outputs into natural-language answers and visual artifacts, illustrating end-to-end, multi-modal reasoning within a single workflow.}
    \label{fig:genie-usecase}
\end{figure*}

To test the end-to-end mechanistic (structure-to-function) reasoning capabilities of Genie-CAT, we use a ferredoxin containing two [4Fe–4S] clusters from {\it Clostridium pasteurianum} (PDB code 1CLF), which is frequently used as a model system for redox-active [4Fe–4S] proteins. This case study illustrates how the agent coordinates retrieval, structural parsing, electrostatic modeling, and ML-based redox prediction to construct mechanistically grounded hypotheses about impact of residue-level modifications in proteins with minimal user intervention.
Figure~\ref{fig:genie-usecase} illustrates the primary step(s) and the results generated by Genie-CAT.\\

\noindent{\bf Step 1: Structural ingestion and descriptor extraction.}
The user query about 1CLF, initiates Genie-CAT to automatically retrieve the structure from PDB and parse with MDAnalysis tools. Genie-CAT identified the two [4Fe–4S] clusters and extracted residue-level descriptors including polarity, solvent accessibility, and atomic distances within a 6~Å cutoff around Fe atoms (Figure ~\ref{fig:genie-usecase}: top-left). Summary tables and heatmaps highlighted modest but noticeable microenvironment differences(Figure ~\ref{fig:genie-usecase}: right): one cluster sits in a slightly more hydrophobic pocket, while the second has a somewhat higher density of nearby polar or charged residues (e.g., Asn, Asp). These differences are consistent with known asymmetries in ferrodoxins.\\

\noindent{\bf Step 2: Electrostatic potential computation.}
Electrostatic potentials were computed using APBS with the Amber ff14SB force field \cite{Maier2015} and in-house Fe–S parameters. Surface visualizations showed anisotropic negative fields around both cubanes, consistent with established electrostatic profiles of ferredoxins. While qualitative, these maps provide physically interpretable signals that contextualize potential redox shifts.\\

\noindent{\bf Step 4: Prediction of the redox potential.} 
From the ESP grid, Genie-CAT derived local electric-field vectors and scalar descriptors for each cluster. These were combined with geometric invariants based on pairwise distances and triple products among cluster atoms, producing a 57-dimensional feature vector. The agent passed this feature vector to the redox-prediction model. The predicted ordering matched expectations for ferredoxins: the cluster in the more hydrophobic environment exhibited a slightly more negative predicted potential (-425 mV), while the cluster with greater local polarity was shifted in the positive direction (-370 mV). Magnitudes were within tens of millivolts of the reported trends for related Fe–S proteins, providing a reasonable internal consistency check.\\

\noindent{\bf Step 5: Correlation and mechanistic interpretation.}
Genie-CAT synthesized these results as sown in Figure ~\ref{fig:genie-usecase}, allowing scientists to form new or validate existing hypothesis: linking local polarity to redox modulation. The agent inferred that the introduction of additional polar residues near the more hydrophobic cluster may stabilize the oxidized states and increase its potential. These predictions are consistent with previously reported trends for bacterial ferredoxins in asymmetric cubane environments \cite{jafari_benchmark_2022}. In general, these suggestions reproduce qualitative design principles for Fe–S proteins and illustrate how Genie-CAT converts structural and electrostatic evidence into actionable hypotheses for downstream evaluation.

% ======================================================================
% DISCUSSION
% ======================================================================
\section{Discussion}
Genie-CAT demonstrates how an agentic LLM can move beyond text-only assistance toward mechanistically grounded hypothesis generation for metalloproteins. 
By integrating literature-grounded reasoning (RAG), structural parsing of PDB files, electrostatic potential calculations, and a redox predictor into a single workflow, the system produces proposals that explicitly tie sequence- and environment-level edits to hypothesized energetic consequences at metal centers. This reduces the gap between idea generation and quantitative analysis, providing users with tractable, testable hypotheses rather than open-ended suggestions.

Genie-CAT's agent significantly reduces the time and expertise barriers for novel hypothesis generation. We estimate that, for a user to perform the sequence of ESP and redox calculations manually, it would take days to weeks depending on expertise. In contrast, Genie-CAT runs such calculations in less than 3 minutes per protein. In addition, the expertise and time savings provided by the RAG component are invaluable, providing reliable scientific grounding by reducing the risk of hallucination. The ferredoxin case study illustrates these benefits in a realistic setting. 

Each component of Genie-CAT addresses a distinct failure mode common in LLM-only workflows. Literature retrieval grounds claims and mitigates hallucination; structural parsing turns text prompts into precise residue- and environment-level contexts; electrostatics provides a physically interpretable field-level signal; and the redox model aggregates geometric and electrostatic descriptors to support ranking and sensitivity analyses. The agent then synthesizes these heterogeneous signals into concise rationales, making assumptions explicit and highlighting the evidentiary basis for each recommendation.

There are significant limitations. First, literature-grounded reasoning depends on corpus coverage and curation; gaps or biases in the ingestion set can propagate to hypotheses. Second, continuum electrostatics (APBS) provides a tractable but approximate description near metal centers, where polarization and quantum effects can be significant. Third, the redox predictor is trained on finite, domain-specific data and may not uniformly generalize across metalloprotein families or unusual coordination motifs; calibration and uncertainty quantification remain active areas of work. Finally, analyses are typically performed on single structures or a small ensemble, and thus may underrepresent conformational heterogeneity.

Despite these caveats, the agentic pattern is extensible. Future iterations can broaden the corpus beyond journal articles to include curated databases and structured reaction/thermodynamic data; incorporate conformational sampling via atomistic molecular dynamics to expose the agent to ensemble-averaged features; and integrate hybrid quantum mechanical/molecular mechanics (QM/MM) or polarizable empirical force-field methods for higher-fidelity local electrostatics when warranted. Beyond Fe--S systems, the same workflow can accommodate heme, Fe--CO, and other inorganic cofactors by extending parameter libraries and descriptors, while tightening the loop with automation and wet-lab feedback to refine both retrieval and predictive components iteratively.

\section{Conclusion}
The results presented here highlight the need for dedicated scientific agents in mechanistic protein design. General-purpose LLM systems excel in broad reasoning, but lack access to the structural, electronic, and thermodynamic evidence required to produce reliable biochemical hypotheses. In contrast, Genie-CAT's domain-specific tooling enables the agent to ground its reasoning in quantifiable physical descriptors, such as electrostatic fields, Fe-S geometry, and symmetry-aware redox features, thereby reducing hallucination and improving mechanistic interpretability. This integration is essential in metalloprotein design, where subtle environment-dependent effects can shift redox thermodynamics by a few tens of millivolts and where high-level textual reasoning alone is insufficient. By combining retrieval, structure, biochemistry, physics, and machine learning within a single orchestrated workflow, a dedicated agent can serve as a scientific assistant that augments expert intuition, accelerates design iteration, and makes previously labor-intensive analyses available through natural language queries.

Genie-CAT links local environment features to predicted redox shifts at [4Fe--4S] clusters, generating concrete residue-level proposals suitable for downstream testing. Looking ahead, expanding corpus coverage, adding ensemble- and QM-aware physics, and integrating uncertainty and experimental feedback will further strengthen the reliability and impact of agent-guided hypothesis generation in protein design.

\section{Acknowledgements}
This research was supported by the Generative AI (GenAI) for Science, Energy, and Security Science \& Technology Investment under the Laboratory Directed Research and Development (LDRD) Program at Pacific Northwest National Laboratory (PNNL) and the U.S. DOE, Office of Science, Office of Basic Energy Sciences, Division of Chemical Sciences, Geosciences, and Biosciences, Physical Biosciences Program under awards FWP 66476. A portion of the research was performed using resources available through Research Computing at PNNL. PNNL is a multiprogram national laboratory operated by Battelle for the Department of Energy under Contract No. DE-AC05-76RLO 1830.

% ======================================================================
% BIBLIOGRAPHY
% ======================================================================
\bibliographystyle{unsrtnat}
\bibliography{citations}

\end{document}